\documentclass[12pt]{article}
\usepackage{fullpage}
\usepackage{filecontents}
\usepackage{graphicx}

\def\3Hei{$^{3}${\rm He}$^{++}$}
\def\He3{$^{3}${\rm He}}
\def\4He{$^{4}${\rm He}}
\def\deg{$^{\circ}\ $}

\usepackage[backend=bibtex]{biblatex}
\bibliography{He3pol}

\begin{document}

\title{Measurement of \3Hei Polarization at 5.3 MeV via Scattering with an Unpolarized \4He Target}
\author{Charles Epstein and Richard Milner \\ Laboratory for Nuclear Science \\ Massachusetts Institute of Technology \\ Cambridge, MA 02139}

\date{July, 2023 (Original Date: March, 2014)}
\maketitle

\subsection*{Preface (2023)}

Since 2012, a BNL-MIT collaboration has worked to develop a polarized $^3$He ion source for the Relativistic Heavy Ion Collider (RHIC) using the existing Electron Beam Ionization Source (EBIS). $^3$He atoms are polarized using optical pumping at high field and injected as neutral atoms into the EBIS.  A critical issue is the demonstration that the polarization of the extracted $^3$He$^{++}$ ions from the source is high.  In 2014, a concept to measure the nuclear polarization was developed in this paper.  Since then, the development of the polarized $^3$He ion source has progressed, the polarimeter is under construction, and the demonstration experiment is expected to take place in the next several years \cite{zelenski}.  

\subsection*{Techincal Note (2014)}

A proposed method of measuring the \He3 polarization post-ionization in EBIS is to observe scattering with a fixed unpolarized \4He target.  In \cite{plattner} it is shown that at several beam energies and center of mass scattering angles, the asymmetry due to polarization of the \He3 beam tends towards 1.  This gives very strong resolving power in determining the beam polarization.  In particular, a beam energy of 5.3 MeV leads to an asymmetry of 1 at 91\deg in the center of mass frame.  This corresponds to an outgoing \He3 at 53.6\deg and 2.66 MeV, and an outgoing recoil \4He at 44.5\deg and 2.64 MeV.  

\begin{figure}
\centering
\includegraphics[width=0.8\textwidth]{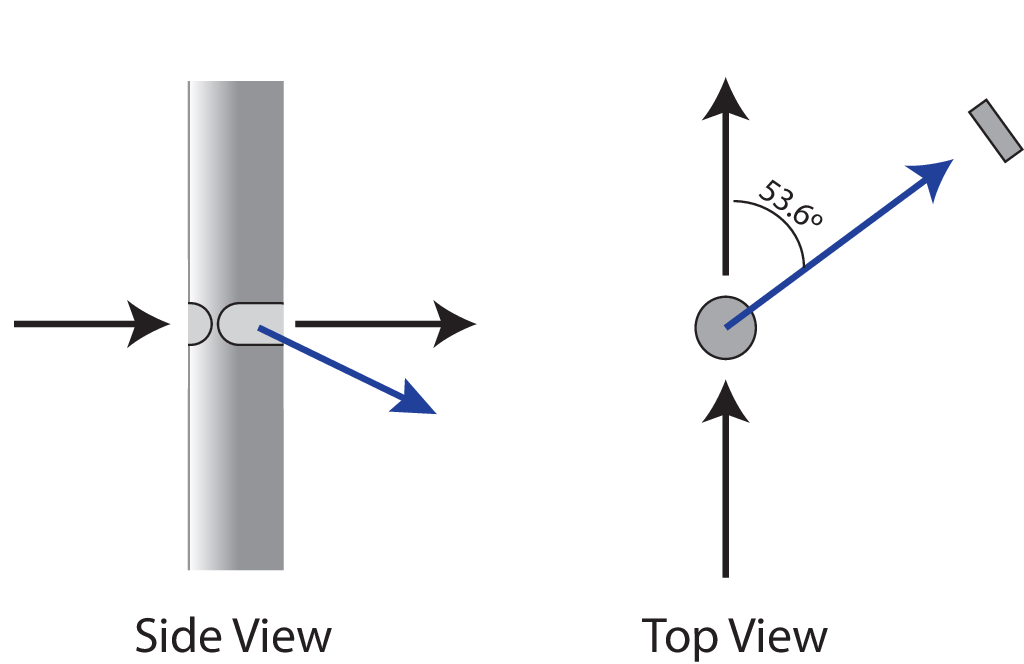}
\caption{Diagram of the target and detector}
\label{diagram}
\end{figure}

In order to measure the cross-section at such kinematics, it is proposed to place a small circular detector of approximately 1 cm diameter at a distance of 10 cm from a small tubular \4He cell operating at a pressure of approximately 5 torr (Fig. \ref{diagram}).  This corresponds to a target thickness of approximately $2\times 10^{17}\ {\rm cm}^{-2}$.  With approximately $10^{12}\ ^3{\rm He}^{++}{\rm s}^{-1}$ extracted from EBIS, this provides a luminosity of ${\cal L} \sim 2\times 10^{2}\ {\rm s}^{-1}{\rm mb}^{-1}$.  In \cite{spiger}, the \4He(\He3,\He3)\4He cross-section was measured.  At 90\deg in the center of mass for \He3 incident at 5.2 MeV, the cross-section peaks at the $^2F_{7/2}$ resonance at 150 mb/sr.  At 5.3 MeV, it is closer to 125 mb/sr.  This indicates that a detector with 1 cm$^2$ area will observe \He3 events at approximately 250 Hz.  

Beam energy losses in the windows and target gas must be considered.  In the cross-section measurements of \cite{spiger}, the \4He target consisted of a closed gas cell with Nickel beam windows of approximate thickness 0.6 $\mu$m.  Using the Bethe-Bloch formula, one estimates that a 5.3 MeV \3Hei ion will lose approximately $7\times10^{-4}$ MeV in traversing two such windows.  This is negligible and corresponds to a power deposition of $\sim0.1$ mW, indicating that a closed cell with such windows is suitable for our polarimeter.  Using the same method, it is estimated that the beam will lose approximately $2\times10^{-4}$ MeV as it travels through the 1 cm, 5 torr target cell.  This is also negligible.  It is proposed that an aluminum pipe of $\sim$1 cm (or $\sim$1/2-inch) diameter, with routed apertures covered with nickel foil, could serve as the target chamber.  

A source of background are events at which \He3 scatters to $\sim 72.8$\deg in the center of mass.  This produces an outgoing recoil \4He at 53.6\deg in the lab frame, the same as the signal \He3.  The energy of this \4He is approximately 1.83 MeV.  Thus, a detector with energy resolution should be able to easily distinguish the 2.66 MeV \He3 from the 1.83 MeV \4He.  A 500 $\mu$m partially-depleted silicon detector available commercially from Canberra (PD Series) or Ortec (A-Series) is proposed.  These are very well-suited to such an application, as they provide good energy resolution and will stop all incoming signal particles, which have a range of $\sim 10\mu{\rm m}$.  

\begin{figure}
\centering
\includegraphics[width=0.75\textwidth]{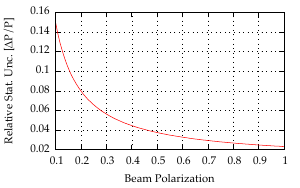}
\caption{Relative statistical uncertainty ($\Delta P/P$) after one minute as a function of beam polarization}
\label{unc}
\end{figure}

In addition, the statistical error on the polarization measurement can be estimated.  Assuming full analyzing power, after $N$ events, this can be expressed as $$\Delta P = \sqrt{\frac{(P+1)(P+3)}{N}}.$$  At 50\% beam polarization, this indicates that one can expect roughly 3.7\% relative statistical uncertainty after one minute of integration time.  A plot of the relative uncertainty as a function of polarization is visible in Fig. (\ref{unc}).  \nocite{hardy}

This work was supported by the Office of Nuclear Physics of the U.S. Department of Energy.

\printbibliography
\end{document}